\documentclass[fleqn,usenatbib]{mnras}

\usepackage{newtxtext,newtxmath}
 
\usepackage[T1]{fontenc}

\DeclareRobustCommand{\VAN}[3]{#2}
\let\VANthebibliography\thebibliography
\def\thebibliography{\DeclareRobustCommand{\VAN}[3]{##3}\VANthebibliography}

\usepackage{graphicx}	
\usepackage{amsmath}	
\usepackage{amssymb}	

\newcommand{\ntwohp}{N$_2$H$^+$}
\newcommand{\codename}{{\tt McFine}}
\newcommand{\tex}{$T_{\rm ex}$}

\graphicspath{{./}{figs/}}

\title[\codename]{\codename: {\tt python}-based Monte-Carlo multi-component hyperfine structure fitting}

\author[T. G. Williams et al.]{
Thomas G. Williams,$^{1}$\thanks{E-mail: thomas.williams@physics.ox.ac.uk (TGW)} and
Elizabeth J. Watkins$^{2}$
\\
$^{1}$Sub-department of Astrophysics, Department of Physics, University of Oxford, Keble Road, Oxford OX1 3RH, UK\\
$^{2}$Jodrell Bank Centre for Astrophysics, Department of Physics and Astronomy, University of Manchester, Oxford Road, Manchester M13 9PL, UK\\
}

\date{Accepted XXX. Received YYY; in original form ZZZ}

\pubyear{2024}

\begin{document}
\label{firstpage}
\pagerange{\pageref{firstpage}--\pageref{lastpage}}
\maketitle

\begin{abstract}
Modelling complex line emission in the interstellar medium (ISM) is a degenerate, high-dimensional problem. Here, we present \codename, a tool for automated multi-component fitting of emission lines with complex hyperfine structure, in a fully automated way. We use Markov chain Monte Carlo (MCMC) to efficiently explore the complex parameter space, allowing for characterising model denegeracies. This tool allows for both local thermodynamic equilibrium (LTE) and radiative-transfer (RT) models. \codename\ can fit individual spectra and data cubes, and for cubes encourage spatial coherence between neighbouring pixels. It is also built to fit the minimum number of distinct components, to avoid overfitting. We have carried out tests on synthetic spectra, where in around 90~per~cent of cases it fits the correct number of components, otherwise slightly fewer components. Typically, \tex\ is overestimated and $\tau$ underestimated, but accurate within the estimated uncertainties. The velocity and line widths are recovered with extremely high accuracy, however. We verify \codename\ by applying to a large Atacama Large Millimeter/submillimeter Array (ALMA) \ntwohp\ mosaic of an high-mass star forming region, G316.75-00.00. We find a similar quality of fit to our synthetic tests, aside from in the active regions forming O-stars, where the assumptions of Gaussian line profiles or LTE may break down. To show the general applicability of this code, we fit CO({\it J} = 2-1) observations of NGC~3627, a nearby star-forming galaxy, again obtaining excellent fit quality. \codename\ provides a fully automated way to analyse rich datasets from interferometric observations, is open source, and {\tt pip}-installable.
\end{abstract}




\section{Introduction}

Spectral line emission is a key tool for diagnosing the physical conditions of the interstellar medium \citep[ISM, e.g.\,][amongst many others]{2008Walter, 2009Leroy, 2016Rigby, 2017Pety, 2019Liu, 2021Leroy}, but extracting useful and accurate physical parameters from observed line intensities is a complex task. If only a single line is measured (for example, CO in most extragalactic studies), we can access the velocity, gas dispersion, and with a conversion factor\footnote{This CO-conversion factor, $\alpha_{\rm CO}$, is either assumed or estimated from some other parameter \citep[primarily the metallicity, see][and references therein]{2013Bolatto}.} the molecular gas mass. With different transitions of the same molecule, we can instead access different phases of the gas \citep[e.g. in density or temperature, see][]{2016Ohashi,2018Penaloza, 2020Puschnig}. Usually, tracing different transitions requires multiple telescope configurations covering different frequency bandwidths, but in some instances, molecules can contain hyperfine structure. This hyperfine structure exhibits as multiple lines at slightly offset frequencies, arising from the splitting of the rotational energy levels by electric quadrupole and magnetic dipole interactions, which is caused by nuclear moments of atoms with non-zero spin. Therefore, we can trace multiple transitions, and thus gas properties such as excitation temperature and optical depth with a single frequency configuration, as modern radio telescopes allow us to resolve these hyperfine lines. These properties offer more insight into the gas conditions than simply measuring the line intensity and velocity since they allow for a detailed study of the local physical conditions of the gas, such as the gas density.

The most straightforward method of calculating the excitation and optical depth is assuming that the kinetic and excitation temperature are equal, known as Local Thermodynamic Equilibrium (LTE), in which case all lines have the same temperature and opacity \citep[e.g.][]{2016Ohashi}. This assumption holds true in high-density regimes (i.e.\ significantly above the critical density of the molecules in question), but breaks down in regions of lower density \citep[see][]{2015MangumShirley}. A next step towards a more physically accurate, but complex model, involves knowledge of the collisional properties of the molecules. In this case, a kinetic temperature is modelled, and the excitation temperature and opacity of each line may be different. Local Velocity Gradient (LVG) methods take advantage of this, as well as the popular {\sc radex} tool \citep[][which we use for the radiative transfer models in this work]{2007vanderTak}. More complex models remove the requirement of local excitation and rather solve the radiative transfer as functions of depth into the cloud \citep[e.g.][]{2002vanZadelhoff}.

The spectra along any given line-of-sight can also show complexity in multiple ways. Many lines will split into multiple hyperfine components, which can be blended together or separated by a small velocity offset. The ISM is continuous and so there may be multiple distinct regions of emission along any given line-of-sight. It is these multiple distinct regions that we will specifically refer to under the term of ``spectral complexity'' throughout this work, though we do acknowledge that many individual spectra can also show complex profiles. Spectral complexity within datasets is ubiquitous \citep[e.g.][]{2015Lindner, 2016Henshaw, 2020Henshaw, 2021Koch, 2024Rigby}, and blindly applying a single component fit to a spectrum with multiple components will at best lead to results that are, on average, correct, and at worst are completely unphysical or misleading \citep{2021Koch}. Defining a number of components per-sightline by hand for the increasing scale of datasets from state-of-the-art radio telescopes is subjective and prohibitively time-consuming, and so this step must be automated along with the fitting of each spectral component.

There are a number of tools available to help automate this task, and for a review of some of these we refer the reader to \cite{2019Riener}. Since that work, there also now exist codes more specific for hyperfine structure, such as MUFASA \citep{2020Chen} and {\tt mwydyn} \citep{2024Rigby}. However, all of these codes often rely on either capping the maximum number of components that can be fit, or require some human interaction to set the number of components. Given the extremely large volume of datasets now regularly obtained with radio instruments, this is becoming increasingly unwieldy. Additionally, the existing tools are highly susceptible to the degeneracies present in the fitting procedure. A tool that can fit hundreds of thousands of spectra with no manual intervention, and fully explores the complex parameter space, is vital for making progress in characterising the ISM.

In this work, we present \codename, a {\tt python}-based tool to automatically fit complex spectral line data within a fully Bayesian/MCMC framework. Although built to fit \ntwohp, this tool can be easily extended to other lines and transitions. \codename\ offers both simple LTE approximation solutions, as well as more complex radiative transfer (RT) solutions based on {\sc radex} calculations. \codename\ is highly parallelised, and is primarily designed to fit data cubes in an efficient manner, although it can also be used on individual spectra. The code is freely available at \url{https://github.com/thomaswilliamsastro/mcfine}, and is {\tt pip}-installable\footnote{\tt pip install mcfine}. Given the much more complex nature of the {\sc radex} fitting, we will focus on LTE in this paper and defer the RT modelling to a future work. Whilst the RT functionality is present in the current version of \codename, it should be considered relatively untested.

The layout of this paper is as follows. In Section \ref{sec:model}, we describe how we build and fit our multi-component spectral model. In Section \ref{sec:synth_tests}, we apply the fitting routine to a number of synthetic spectra to test the efficacy of the fitting approach. In Section \ref{sec:real_app}, we apply \codename\ to real data, showing the quality of the fits achieved. Finally, we conclude in Section \ref{sec:conclusions}.

\section{Fitting the hyperfine model}\label{sec:model}

\subsection{Model setup}\label{sec:model_setup}

\subsubsection{LTE Case}

In building a hyperfine model, we require an optical depth, $\tau_i$ at the centre of each hyperfine component and an excitation temperature for each component, \tex. In the LTE case, \tex\ is the same for all hyperfine components, and the line strengths are fixed so that the sum of each individual $\tau_i$ is a free parameter in the model.

With an excitation temperature and optical depth for each line, we then build a spectrum. We make a number of assumptions at this point, mainly that the opacity is Gaussian in velocity, and that the line width is the same for each hyperfine line\footnote{This is the same as the GILDAS implementation; \url{https://www.iram.fr/IRAMFR/GILDAS}.}. The opacity as a function of velocity is 
\begin{equation}\label{eq:opacity_gaussian}
    \tau_{v, i} = \tau_i \times \exp-\frac{v - (v_i + v_{\rm cen})}{2\sigma^2},
\end{equation}
where $v_i$ is the offset velocity with respect to the velocity of the central line $v_{\rm cen}$. The observed intensity is given by
\begin{equation}\label{eq:temp_sum}
    T = \sum_i (1 - \exp[\tau_{v, i}]) \times (T_{i, \rm ex} - T_{\rm BG}),
\end{equation}
where $T_{\rm BG}$ is the background temperature, typically 2.73~K, the value for the cosmic microwave background. At very low and high values for the opacity ($\tau \lesssim0.1$, $\tau \gtrsim30$), the $\tau$ term will drop out and the problem becomes degenerate, and (in the low limit, the line ratios do not depend on optical depth and in the high limit the exponential becomes very small), so we add a bound at these points. For each distinct line we therefore have four free parameters in the LTE case -- \tex, $\tau$, $v_{\rm cen}$ and $\sigma$. We fit $\tau$ in natural log-space, which allows the algorithm to more efficiently explore the parameter space\footnote{\codename\ allows the user to also use a linear $\tau$, but in our testing we found this often led to the algorithm becoming stuck in certain regions of parameter space given the large dynamic range in this parameter compared to the others.}. This method is described in more detail in Section \ref{sec:model_fit}. Given that our line profiles are assumed to be Gaussian, this means we are not able to model, for example, infall, outflow, or self-absorption, which cause deviations from Gaussianity. This is likely to manifest as a poorer fit to the data, which can be explored using the statistics that \codename\ outputs (see Sect. \ref{sec:synth_tests}).

\subsubsection{RT Case}

For the {\sc radex} RT case, \tex\ can vary between hyperfine components and so we generate a grid of models based on the free parameters within {\sc radex} (the kinetic temperature $T_{\rm kin}$, H$_2$ density $n_{\rm H_2}$, the column density of the molecule $N_{\rm col}$, and the line width $\sigma$). From this grid, we extract \tex\ and $\tau_i$ for each component in the hyperfine structure, interpolating between grid values where necessary (a requirement to limit the computational cost of these calculations). To generate the grid, we use {\sc ndradex}\footnote{\url{https://github.com/astropenguin/ndradex}}, a wrapper around {\sc radex}, which we have edited to fix an issue with parsing values from lines with identical frequencies\footnote{\url{https://github.com/thomaswilliamsastro/ndradex}}.

Overall, we have an extra free parameter bringing the total number to five: $T_{\rm kin}$, $N_{\rm col}$, $n_{\rm H_2}$, $v_{\rm cen}$ and $\sigma$. We enter these values into a pre-generated {\sc radex} grid, which computes a \tex\ and $\tau$ for each component. These, along with $v_{\rm cen}$ and $\sigma$, are then used in Eqs. (\ref{eq:opacity_gaussian}) and (\ref{eq:temp_sum}). Here, the column and number densities are both fit in log-space as they have a much larger dynamic range than the other free parameters.

\subsection{Model fitting}\label{sec:model_fit}

\begin{figure*}
\includegraphics[width=\textwidth]{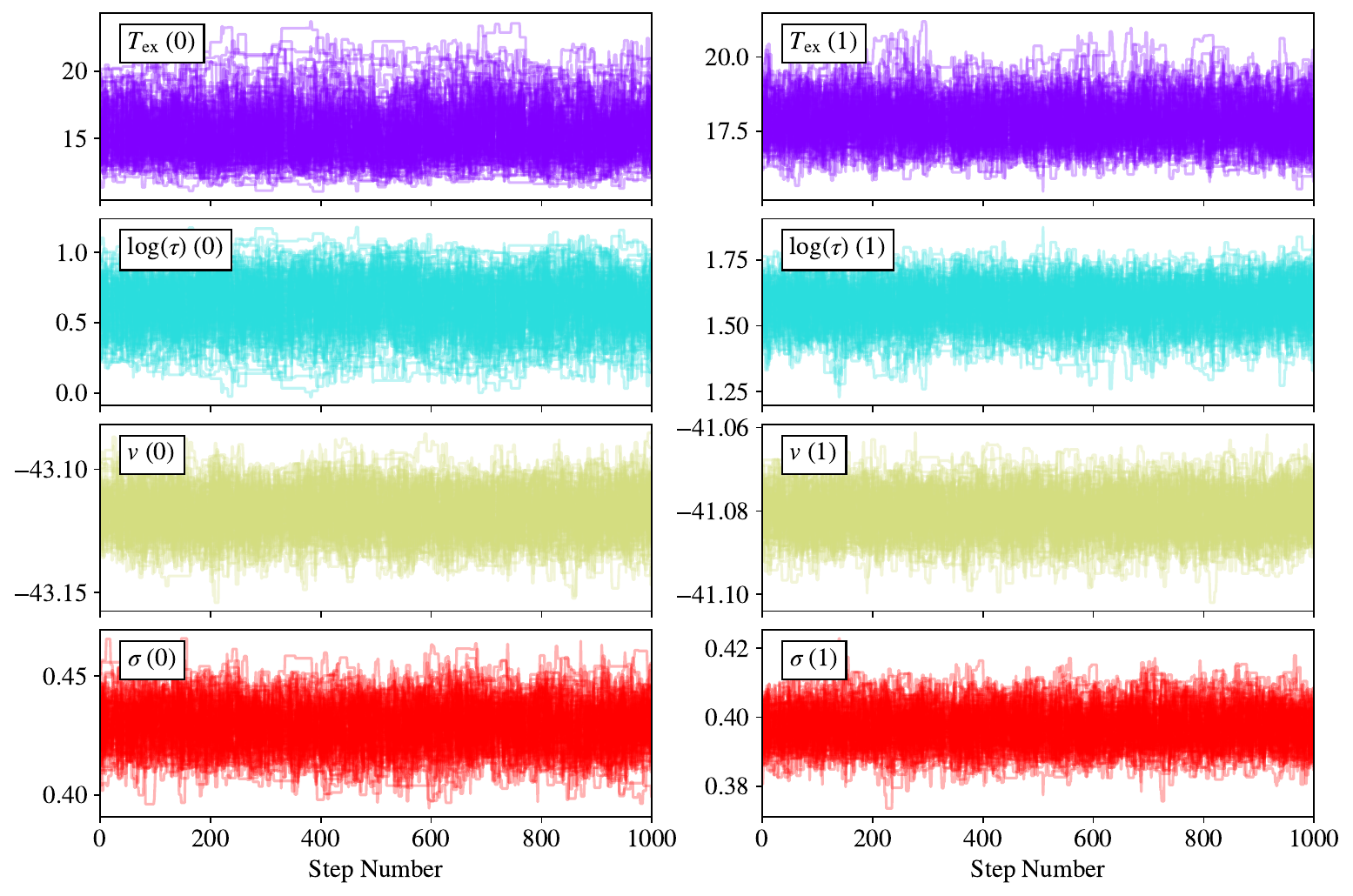}
\caption{Step plot for one of our synthetic spectra tests. We can see the individual walkers are well mixed, and have converged to a stable value for each parameter.} 
\label{fig:step}
\end{figure*}

\begin{figure}
\includegraphics[width=\columnwidth]{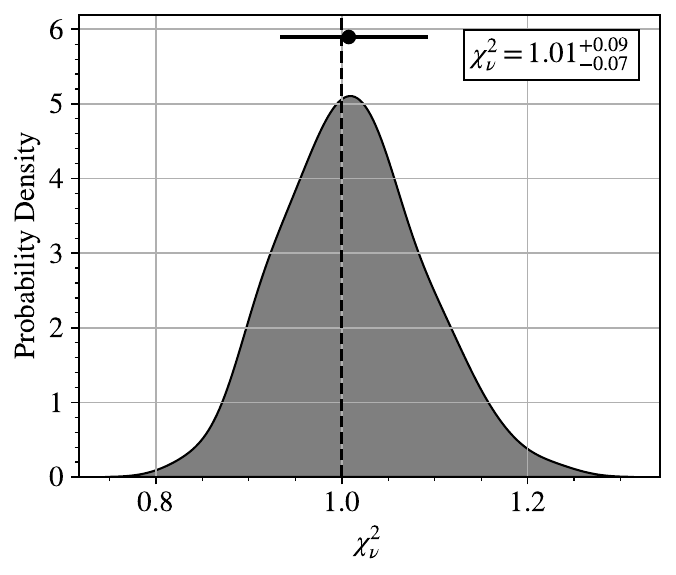}
\caption{Distribution of reduced chi-square values for our synthetic spectra tests. The black point and error-bar show the median, and the 16$^{\rm th}$ and 84$^{\rm th}$ percentiles of the distribution (also given in the text box). $\chi_\nu^2 = 1$ is shown as a dashed vertical line.} 
\label{fig:synth_red_chi_square}
\end{figure}

\begin{figure}
\includegraphics[width=\columnwidth]{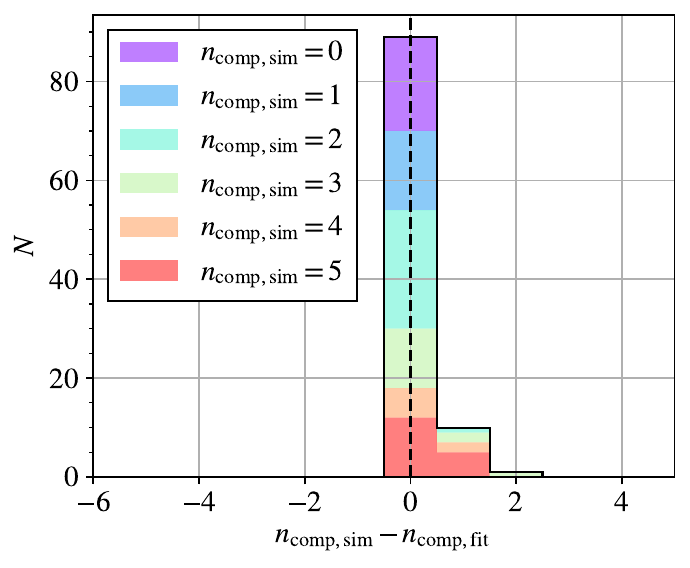}
\caption{Distribution of input minus fitted components for our 100 generated spectra. We colour the histogram by the number of input components, and indicate where the number of fitted components is equal to the number of input components as a dash-black, vertical line.} 
\label{fig:n_comp_comparison}
\end{figure}

\begin{figure*}
\includegraphics[width=\textwidth]{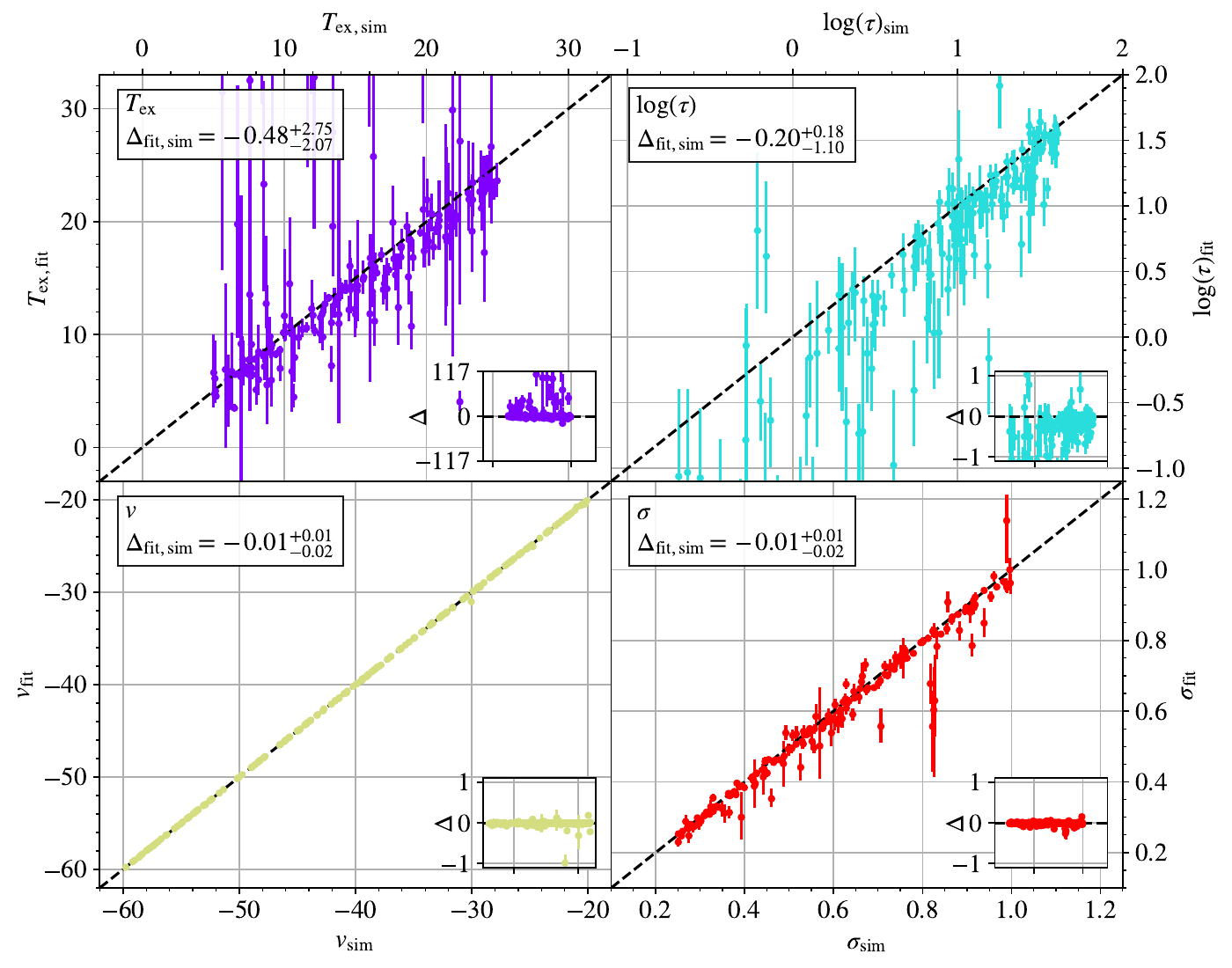}
\caption{Comparison between input and fitted parameters for the models where the correct number of parameters are fitted. The points are coloured by parameter to match those of Figure \ref{fig:param_comparison_sigma}. In each case, the abscissa shows the input parameter value, and the ordinate the fitted parameter. In each subplot we also show the median and offset values from the truth in a text box in the top left, and the 1:1 line is shown as a dashed black line. The small inset plot shows the residual, having removed this 1:1 line, to make any deviations clearer.} 
\label{fig:param_comparison}
\end{figure*}

We perform our fitting in a fully Bayesian framework using the MCMC sampler {\sc emcee}\footnote{\url{https://emcee.readthedocs.io/en/stable/}}. Unlike least-squares fitting, MCMC can robustly account for covariance between parameters (particularly important for excitation temperature and opacity). It can also account for asymmetric uncertainties, which we find to be typical in our fits. To fit the data, we build up a steadily more complex model starting from zero components (i.e. a flat line). The (ln-)likelihood function is given by
\begin{equation}
    \ln\mathcal{L} = -0.5 \sum_i \left(\frac{T_{\rm obs} - T_{\rm model}}{\sigma_{\rm obs}}\right)^2,
\end{equation}
where $T_{\rm obs}$ and $T_{\rm model}$ are the observed and modelled intensity, respectively, and $\sigma_{\rm obs}$ is the noise in the spectrum. We use 100 walkers (with starting positions perturbed around the initial guess) for each fit. Each walker takes 250 steps of burn-in and then 1000 steps for the main run, and both of these are variables that users can alter as desired. We chose these default parameters from inspecting where the walks converged by eye (i.e.\ where the walkers in the step plot appear well-mixed; see Figure \ref{fig:step} for an example). We will implement a more rigorous convergence criterion in a future release. We calculate a maximum likelihood from the median parameters of this fit, discarding the first half of the steps as the walkers explore the parameter space. Throughout this fitting, we make use of {\tt python}'s multiprocessing capabilities either by sending walkers to separate cores in the case of a single spectrum fit, or by passing each spaxel (i.e.\ the spectrum corresponding to a single $x-y$ position in a cube) fit to a separate core in the case of a data cube.

The MCMC walkers require an initial guess, and for this we find a least-squares fit using {\tt lmfit}. The initial guesses are configurable -- we recommend setting an \tex\ similar to the peak temperature in the cube, a $\ln(\tau)$ of 0, a velocity similar to the systemic velocity of the object and a velocity dispersion of $1~{\rm km~s}^{-1}$, for most cases. However, as long as these parameters are reasonable then {\tt lmfit} should be able to find an acceptable guess to initialise the MCMC. By default, we use the global {\tt basinhopping} method (although this can be tweaked as users require) as we find it to produce good initial estimates even into a high-dimensional parameter space. Note that the exact details of the least-squares fit are not vital, but should simply provide a good initial guess for the parameters. If the initial guesses are far from the `true' parameters, the MCMC will typically not converge and the fit will be poor. This is especially true for more complex fits as the parameter space becomes increasingly more complicated. This is a general problem in fitting data, and not unique to our problem. 

After fitting for one component, we gradually increase the complexity by adding components one at a time. When we increase the number of components, we {\it do not} use any of the parameters from the fewer component fits to guide our procedure. Each fit is done completely independently, and then compared at the end. We discriminate between the more complex and previous model using the the Bayesian information criterion (BIC). This is given by
\begin{equation}
    {\rm BIC} = k \ln(n) - 2\ln(\mathcal{L}),
\end{equation}
where $k$ is the number of fitted parameters, and $n$ the number of points that are fit to. A change in BIC, $\Delta {\rm BIC} > 10$ is typically used as indicating a strong preference for one model over another \citep[see, e.g.][]{2021Koch}. We continue to add components until the change in BIC is less than 10, at which point we stop adding components and take the less complex fit as our end point.

Although this procedure does well at identifying the majority of the spectral features, the observational noise is sometimes modelled as a faint, wide spectral feature. We therefore take an additional `top-down' approach where we remove components from faintest to strongest (in terms of integrated intensity). We then fit using the median value of each fitted parameter from the MCMC run as a starting point, and calculate the $\Delta {\rm BIC}$ again. This assumes that the faintest component is most likely to be noise, not necessarily that it is the ``worst'' fit -- in some cases, these components are indeed real, and the comparison will retain such features. If the model with fewer components is not {\it strongly} preferred (i.e.\ $\Delta {\rm BIC} < 10$), then we stick with our original, more complex model.

Finally, for spectral cubes, neighbouring spaxels are not independent (both because the pixel size is selected to oversample the beam, and the gas properties should be at least somewhat correlated over short distances). We therefore have an option to compare the fit parameters to the neighbouring eight pixels. If any of these are strongly preferred by the BIC, we take the most preferred model. We loop over pixels in RA and Dec, first forwards and then backwards. This ends up replacing less than one~per~cent of pixels (see Sect. \ref{sec:real_app}) in our tests on real data sets. Note that at this step we do not perform any refitting, but rather compare the spectrum from the median of each fitted parameter to the observed spectrum in each spaxel.

\section{Tests with synthetic spectra}\label{sec:synth_tests}

\begin{figure}
\includegraphics[width=\columnwidth]{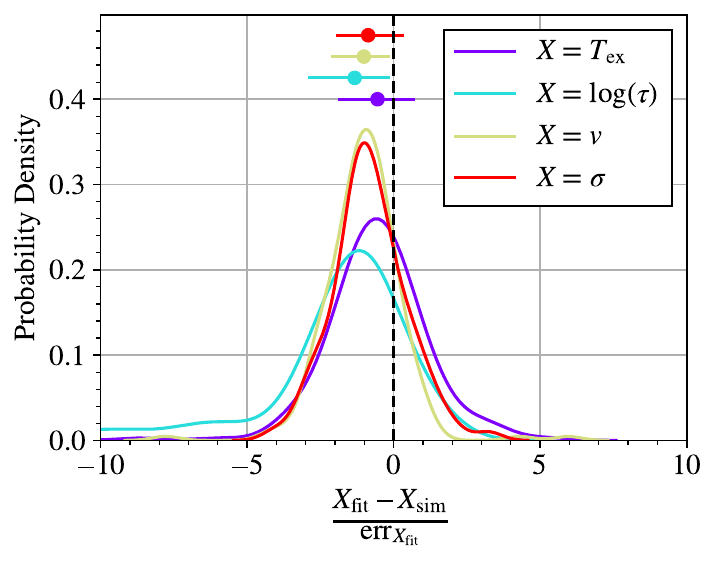}
\caption{Difference between fitted and input parameters over our synthetic models, normalised by the associated errors. Lines are coloured by the particular parameter, which are given in the Figure legend. The vertical, dashed black line indicates where the fitted and input parameter are equal.} 
\label{fig:param_comparison_sigma}
\end{figure}

\begin{table}
	\centering
	\caption{Bounds for randomly drawn parameters in our synthetic tests. $n_{\rm comp}$ is number of components, \tex\ the excitation temperature, $\tau$ the optical depth, $v_i$ the offset from the $-40~{\rm km~s^{-1}}$ centre of the spectrum, and $\sigma$ the line width.}
        \label{tab:synth_params}
	\begin{tabular}{lr}
		\hline
		Parameter & Bounds \\
		\hline
            $n_{\rm comp}$ & [0, 5] \\ 
            \hline
            \tex\ (K) & [5, 25] \\
            $\tau$ & [0.5, 5] \\
            $v_i$ (${\rm km~s^{-1}}$) & [$-20$, $20$] \\
            $\sigma$ (${\rm km~s^{-1}}$) & [$0.25$, $1$] \\
            \hline
	\end{tabular}
\end{table}

To test the efficacy of our modelling approach, we generate a selection of models and apply \codename. We will focus on the LTE approach in this work for simplicity. We generate a number of synthetic spectra with a velocity resolution of $0.2~{\rm km~s^{-1}}$, Gaussian noise with a standard deviation of 0.1~K and a velocity range between $-80$ and $0~{\rm km~s^{-1}}$. This spectral resolution, range, and noise level are all approximately matched to the real ALMA data we will test later . However, we increase the velocity range to account for the potential wide separation in components for these tests, to more appropriately match those seen between large-scale dynamic features in galaxies \citep[e.g.][]{2017Beuther}. We also include a 5~per~cent `calibration' uncertainty, to avoid biasing high $\chi^2_\nu$ values for bright spectra (see Sect. \ref{sec:real_app}). The number of components and parameter bounds for each component are given in Table \ref{tab:synth_params}. 
We generate and fit a total of 100 models to explore this parameter space and the limitations of the fitting. We show some of these fits in Appendix \ref{app:synth_fits}.

Our first metric in this comparison is the reduced chi-square, $\chi_\nu^2$, which we calculate from the $\chi^2$ value divided by the number of free parameters ($n - k$). The distribution of this is shown in Fig. \ref{fig:synth_red_chi_square}, and it shows that the models perform well with a value $\chi_\nu^2 = 1.01_{-0.07}^{+0.09}$, consistent with the expected $\chi_\nu^2 = 1$ for a perfectly fit model. The $\chi_\nu^2$ distribution is approximately Gaussian, indicating no particular bias towards under- or overfitting.

Secondly, we investigate the number of components recovered in our synthetic tests, and we summarise this in Fig \ref{fig:n_comp_comparison}. Around two-thirds of the models recover the correct number of components, with no fits including more components than input. This indicates that \codename\ successfully avoids overfitting spectra. Some of the fits have fewer components than input, and when we examined these found them to be models with highly blended components. This causes the model to preferentially fit a single, brighter component than attempting to split into multiple components. Reassuringly, we find no broad components fitted to the noise. In general, the number of components fit by \codename\ should be considered a lower limit to the true spectral complexity. 

For the 89 models that fit the correct number of parameters, we compare the fitted parameters for each component to the input model parameters, and we show these in in Figure \ref{fig:param_comparison}. We see a good correspondence between the input and fitted parameters, particularly with the velocity and velocity width. Typically, \tex\ tends to scatter above the one-to-one line and $\tau$ scatters low, and this is driven by the degeneracy between these two parameters. We also show the distributions of these parameters normalised by their associated errors in Figure \ref{fig:param_comparison_sigma}. We see that despite some systematic underestimating of parameters, the fitted parameters are within one standard deviation of the true value in $\simeq$50~per~cent of cases (we should expect this number to be 68~per~cent, which is similar). The exception to this is $\tau$, which is very slightly outside the 1-$\sigma$ range.

These synthetic tests show that \codename\ typically produces very good fits to input data, and so we are confident in our fitting procedure going forward and apply it to real data.

\section{Application to ALMA data}\label{sec:real_app}

\begin{figure}
\includegraphics[width=\columnwidth]{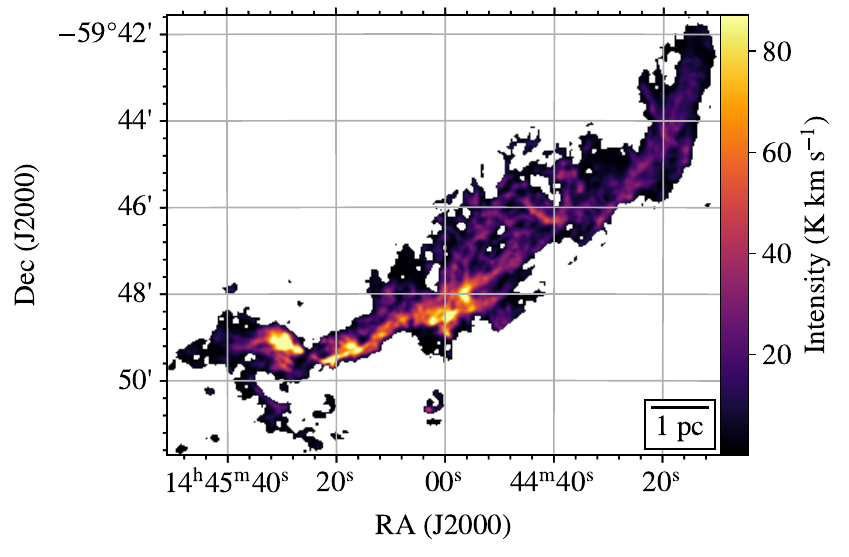}
\caption{Moment 0 (integrated intensity) map of G316.75. A physical scalebar is shown in the bottom right corner.} 
\label{fig:g316_75_mom0}
\end{figure}

\begin{figure*}
\includegraphics[width=\textwidth]{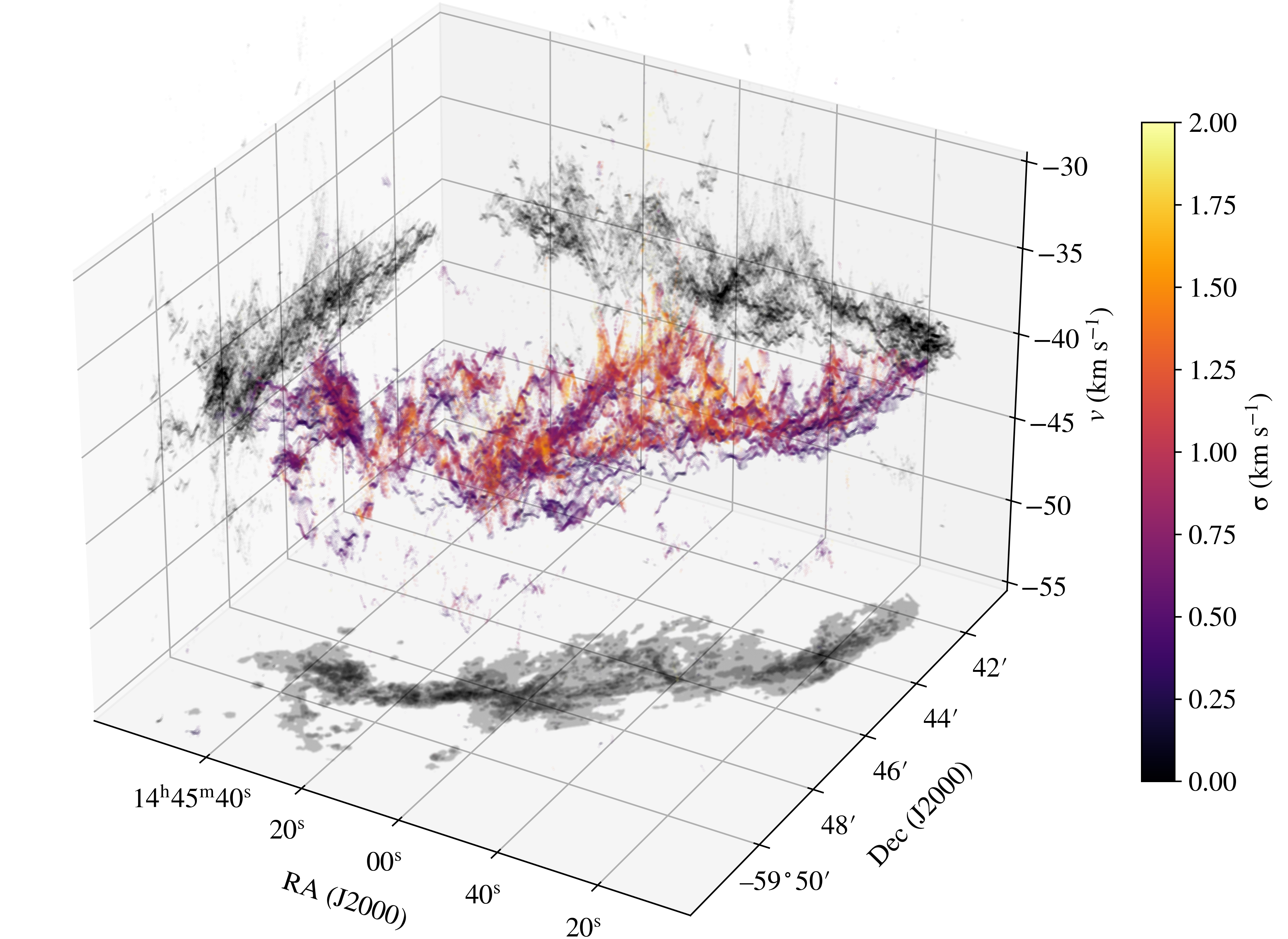}
\caption{Position-position-velocity scatter plot of the \codename\ fits to the G316.75 ALMA data. Points are coloured by the measured velocity dispersion and the overall transparency is intensity-weighted. The projections along each axis are shown in greyscale, so that in the position-position plane, this is equivalent to the number of components along the line-of-sight.} 
\label{fig:g316_75_3d}
\end{figure*}

\begin{figure*}
\includegraphics[width=\textwidth]{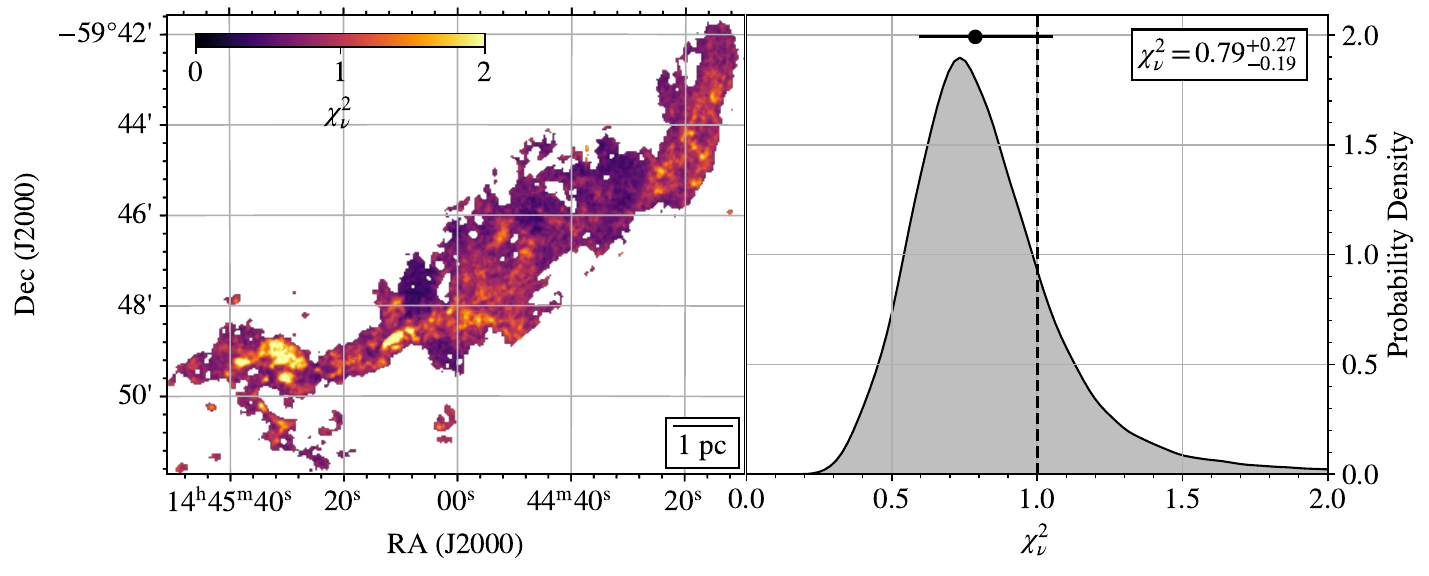}
\caption{{\it Left:} $\chi_\nu^2$ map for each fit within the mask of G316.75. {\it Right}: Distribution of $\chi_\nu^2$ values. The black point and errorbar shows the median value and $16^{\rm th}$/$84^{\rm th}$ percentiles (also given in the box in the top right). A vertical dashed black line shows $\chi_\nu^2=1$.} 
\label{fig:g316_75_chisq}
\end{figure*}

\begin{figure*}
\includegraphics[width=\textwidth]{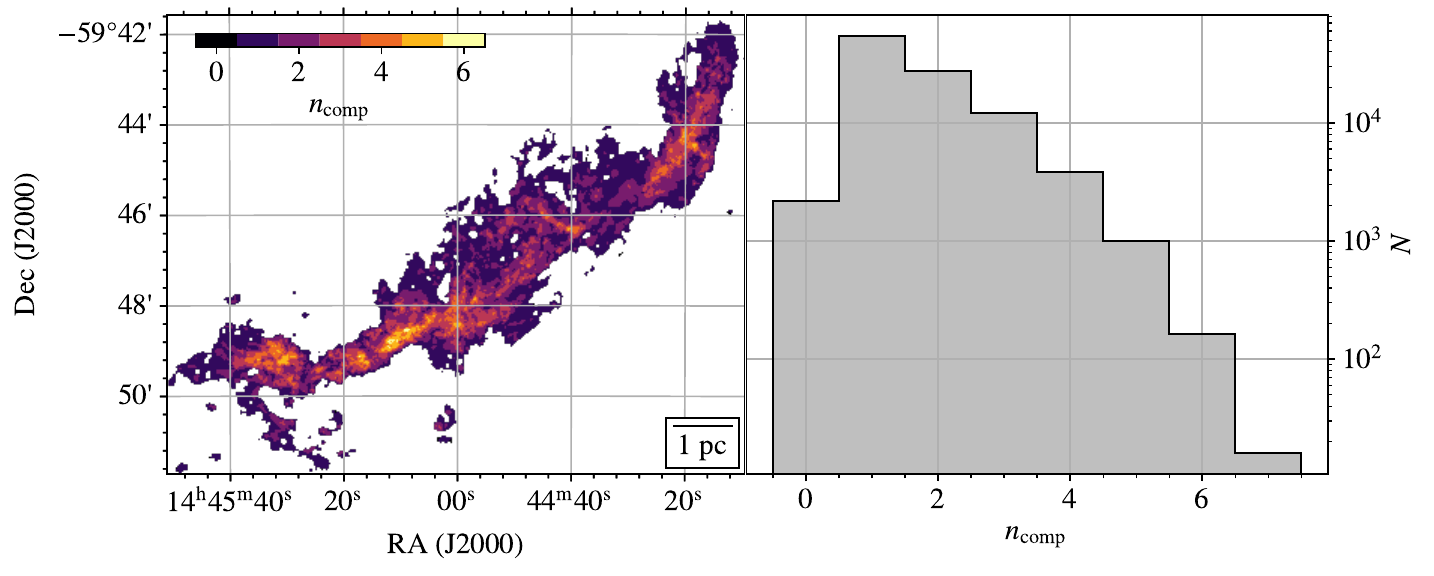}
\caption{{\it Left:} $n_{\rm comp}$ map for each fit within the mask of G316.75. {\it Right}: Histogram of $n_{\rm comp}$ values.} 
\label{fig:g316_75_n_comp}
\end{figure*}

\begin{figure}
\includegraphics[width=\columnwidth]{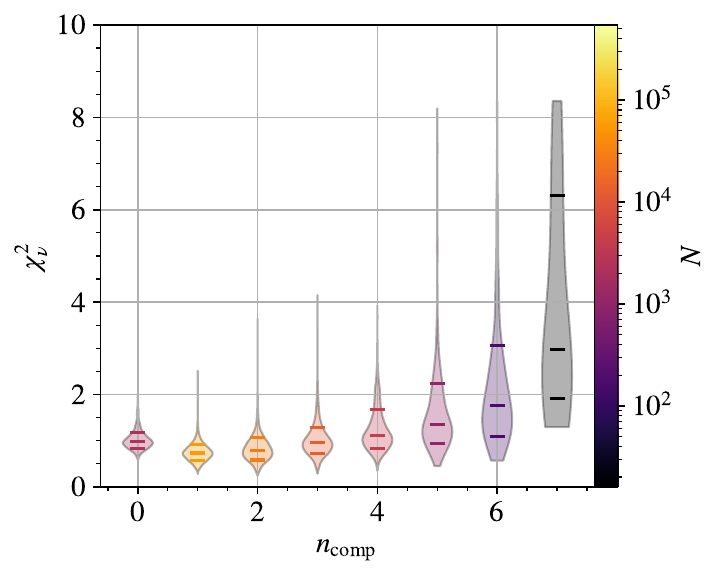}
\caption{Violin plots showing $\chi_\nu^2$ distribution with fitted number of components, $n_{\rm comp}$. Each violin is coloured by the number of points fitted with that number of components, and in each violin we show the median value, along with the $16^{\rm th}$ and $84^{\rm th}$ percentiles of the distribution.} 
\label{fig:g316_75_n_comp_chisq}
\end{figure}

\begin{figure}
\includegraphics[width=\columnwidth]{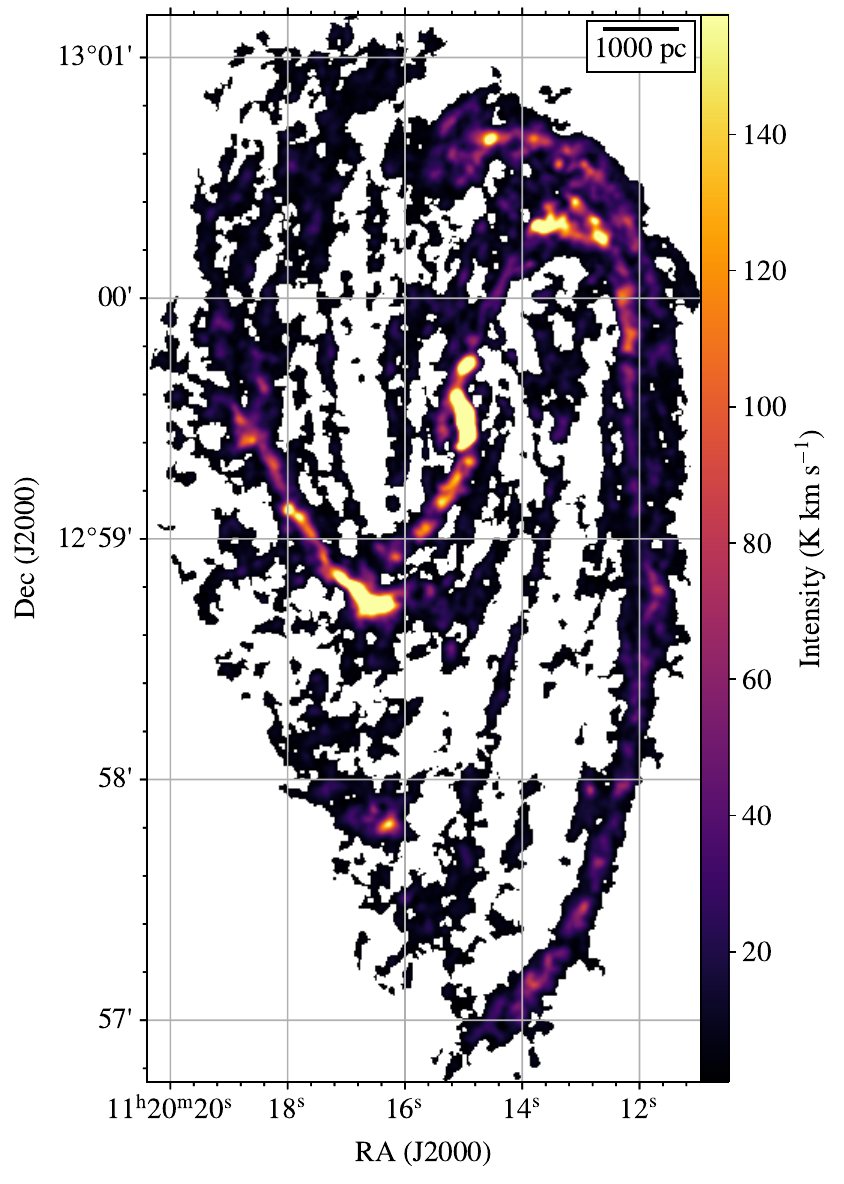}
\caption{Moment 0 (integrated intensity) map of NGC~3627. A physical scalebar is shown in the upper right corner.} 
\label{fig:ngc3627_mom0}
\end{figure}

\begin{figure*}
\includegraphics[width=.8\textwidth]{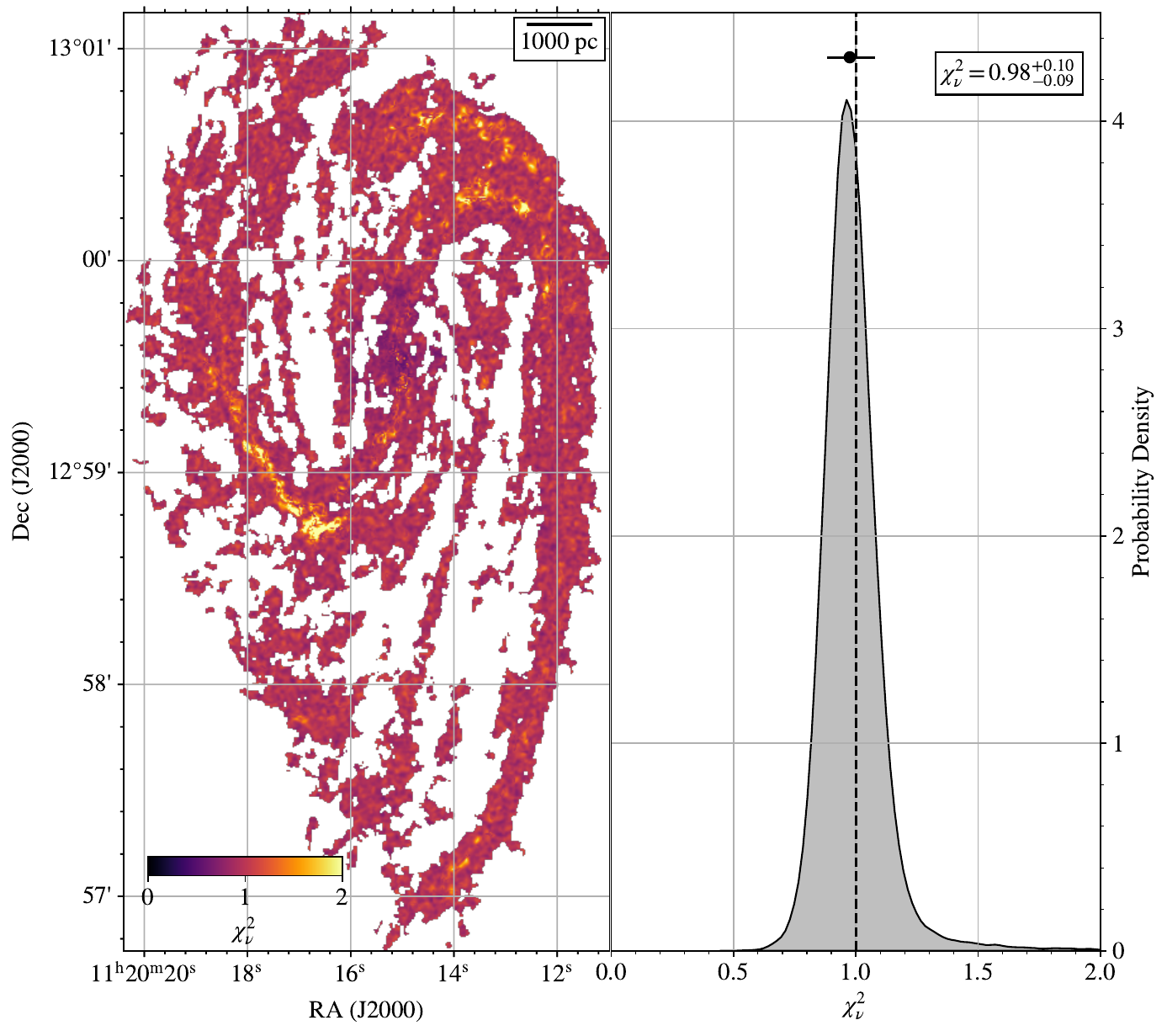}
\caption{As Figure \ref{fig:g316_75_chisq}, but for NGC~3627. To highlight the much more uniform $\chi_\nu^2$, we use the same colourbar scale as in Figure \ref{fig:g316_75_chisq}.} 
\label{fig:ngc3627_chisq}
\end{figure*}

\begin{figure*}
\includegraphics[width=.8\textwidth]{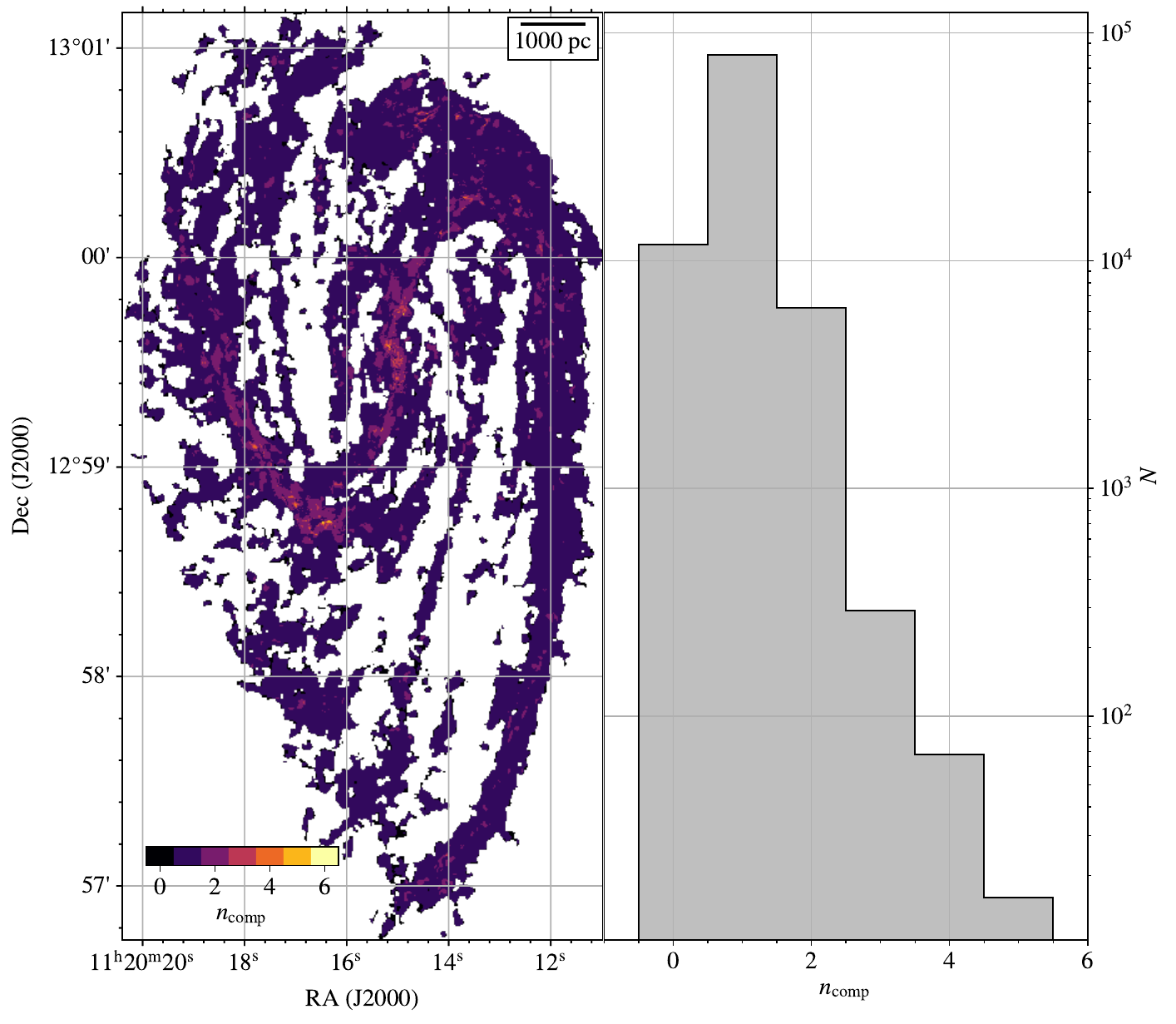}
\caption{As Figure \ref{fig:g316_75_n_comp}, but for NGC~3627.} 
\label{fig:ngc3627_n_comp}
\end{figure*}

\begin{figure*}
\includegraphics[width=\textwidth]{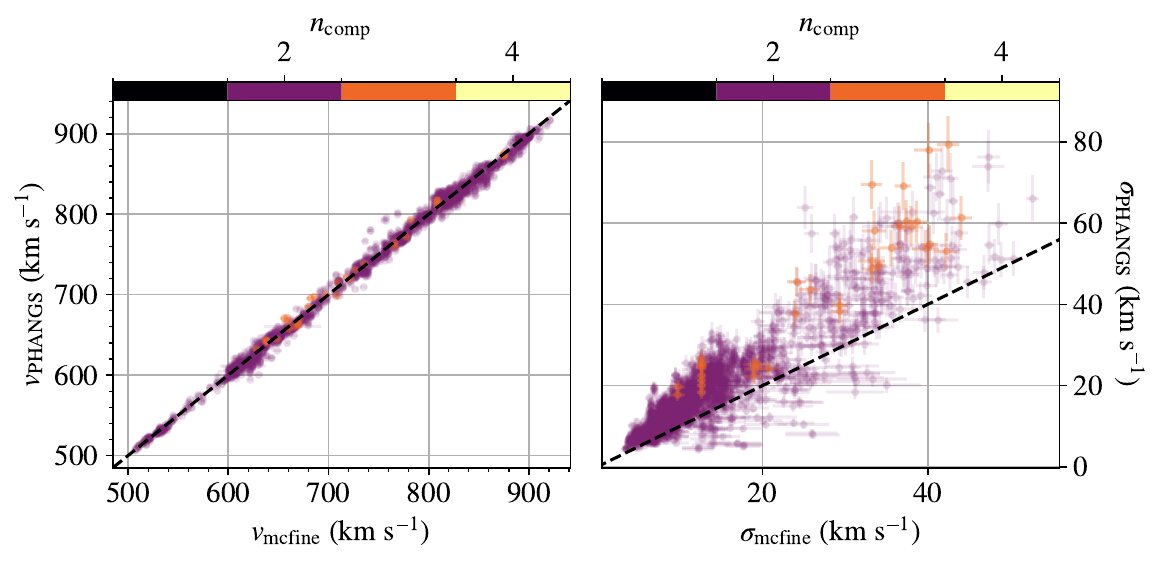}
\caption{{\it Left}: Comparison between mom1 velocities from PHANGS and the peak temperature-weighted average of \codename\ fitted velocity components along each line-of sight. The one-to-one correspondence is shown as a black dashed line. {\it Right}: Comparison between equivalent width from PHANGS and peak temperature-weighted average of \codename\ fitted velocity dispersions for each component along each line-of-sight. Again, the one-to-one correspondence is shown as a black dashed line. Points are coloured by the number of fitted \codename\ components.}
\label{fig:mcfine_phangs_comparison}
\end{figure*}

\begin{figure}
\includegraphics[width=\columnwidth]{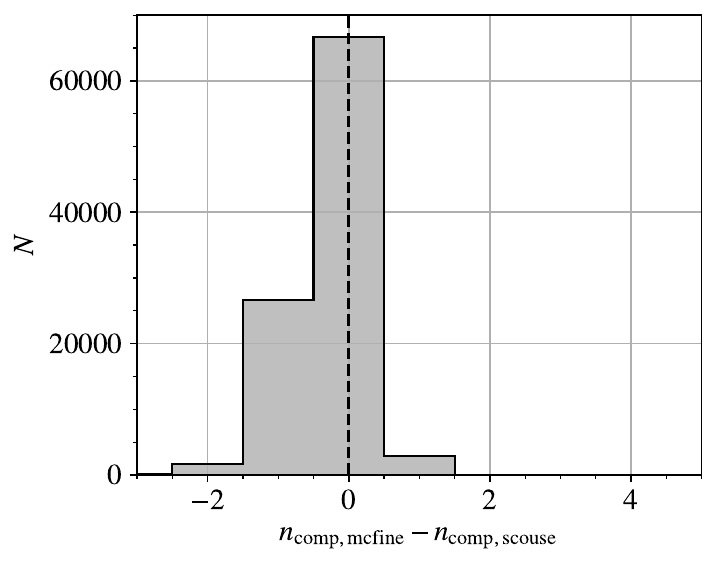}
\caption{Comparison between number of components fitted by \codename\ and {\tt scouse}. The vertical dashed black lines indicates where the same number of components are fit by both codes. Increasingly positive (negative) values indicate relatively more components fitted by \codename\ ({\tt scouse}).} 
\label{fig:mcfine_scouse_comparison}
\end{figure}

In this Section, we provide two benchmarks of the code on ALMA data. The first is G316.75, a region in the Milky Way (MW) observed in the \ntwohp\ line, and which we have focussed \codename\ on fitting. Secondly, to demonstrate the versatility of the fitting code, we apply \codename\ to CO({\it J}=2-1) observations of NGC~3627 (a nearby star-forming galaxy).

\subsection{G316.75}\label{sec:fit_g316}

For the first real-world benchmark for \codename, we apply the fitting to an \ntwohp({\it J}=1-0) mosaic of G316.75-00.00 (G316.75 hereafter), which has been fully observed with ALMA using the 12m, 7m, and total power (TP) arrays at $\sim4$\arcsec\ resolution, equivalent to a physical resolution of 0.036~pc (project codes 2013.1.01029.S and 2018.1.01851.S). These observations comprise three large 12m, 7m, and total power (TP) observations to be sensitive to emission at all angular scales. The integrated intensity map for this mosaic is shown in Figure~\ref{fig:g316_75_mom0}. G316.75-00 is a dense, high-mass, star-forming molecular filament located 2.69$\pm$0.45~kpc away in the Galactic plane \citep{2019watkins}. The filament contains $\sim$20,000~M$_\odot$ of molecular gas over a 14~pc length and exhibits a range of star-forming features in different evolutionary stages. More specifically, the southern half of the filament is dominated by an active H{\sc ii} region powered by 2--4 O-stars. The northern half -- which is just as massive -- is a quiescent infra-red dark cloud (IRDC) containing large star-forming clumps and has hints of a more intricate sub-filament network fuelling their growth in dust emission.

Since \cite{2019watkins} found that G316.75 has overlapping velocity components using lower resolution (0.5~pc) \ntwohp\ observations \citep{2011Foster, 2013Foster, 2013Jackson}, it is an excellent test case for \codename. Due to the range of evolutionary states present, we expect complex, non-Gaussian velocity motions throughout the filament, including outflows and inflows, as well as conditions that might challenge the thermalised assumption of LTE modelling. These features will allow us to evaluate \codename's ability to extract distinct velocity structure in extreme, non-ideal environments, and gauge the underlying limitations of our models.

The full details of the data reduction will be given in Watkins et al. (in preparation), but in brief, we image each mosaic separately, as they have somewhat different noise parameters and observing conditions. Broadly following the strategy of \citep{2021aLeroy}, we combine the 12m and 7m data into a single measurement set (MS) per mosaic, but instead of using {\tt tclean} on the interferometric data and then {\tt feathering} the TP after deconvolution, we use the experimental {\tt sdintimaging} routine to include the TP imaging in the deconvolution. The full details of this algorithm are given in \cite{2019Rau}, and we find this correctly recovers the total flux compared to the TP map, whereas using the more typical {\tt tclean} plus {\tt feather} method underestimates the flux by up to 30~per~cent in the brightest regions. We note that this is typically not an issue in extragalactic observations (as deviations from {\tt tclean/feather} and {\tt sdintimaging} only start to arise at a beam temperature per-channel of around 2~K), but will likely be more problematic in Milky Way targets and extremely bright regions of extragalactic objects. We first perform a shallow multi-scale clean down to $4\sigma$, followed by a deep single-scale clean to $1\sigma$, or a relative flux change in the model flux of less than 0.1~per~cent. Finally, we linearly combine the mosaics based on their noise properties, and homogenise the beam to a common, round beam.

To define a mask that will contain almost all emission, we take the strict mask produced by the reduction pipeline. The details of the mask are given in section 7.3 of \cite{2021aLeroy}, but briefly these grow 4-$\sigma$ peaks of emission down to a 2-$\sigma$ level, thus capturing most emission but maintaining a low false-positive rate. We take any spaxel that has significant emission anywhere down the line of sight as defined by the mask cube. This leads to a total of $\sim$100,000~pixels. We run \codename\ on this cube on a compute server, using 96 CPU cores. This server achieves around 1,000 fits per hour, although individual fit time is highly dependent on the model complexity (i.e. a two component model will take around a factor of two times longer to fit, as it must first fit a single component model). We then loop forth and back over RA and Dec, comparing the models of neighbouring pixels to encourage spatial coherence. We find this replaces $\sim$700 fits, less than one~per~cent. This would indicate that in the vast majority of cases we successfully find the best fit in our first pass. In total, this fitting takes around 120~hours. This means that each fit takes around 5~mins on average on a single CPU core, and so for large data volumes we highly recommend using a high-performance compute cluster.

A position-position-velocity (PPV) rendering of the fits is shown in Figure \ref{fig:g316_75_3d}. A rich variety of structure is seen that would be missed in traditional moment analysis; there is complex structure along the ridge of the IRDC, but also a significant amount of emission above and below the filament. This off-filament gas often has a higher velocity dispersion, indicative of infall or gas ejection \citep[e.g.][]{2010KlessenHennebelle, 2019watkins}. It is also clear that a full 3-dimensional treatment is necessary to build up associations of coherent velocity structures -- any single plane of this PPV space misses crucial information.

We will defer a full exploration of these fits to Watkins et al. (in preparation), but as a first demonstration of the fit quality we show the fitted $\chi_\nu^2$ map and distribution in Fig. \ref{fig:g316_75_chisq}. Note that in calculating $\chi_\nu^2$, we include a 5~per~cent systematic error added in quadrature to the RMS uncertainty, similar to the scatter in flux calibration seen in \cite{2021aLeroy}. We include this to account for the fact that even if the fit quality looks good in the brightest regions, the very small RMS uncertainties lead to extremely large values of $\chi_\nu^2$. In general, we produce extremely good fits to the data, but there are some exceptions (see Appendix \ref{app:alma_fits}). Particularly around the south-east, there is a region with high $\chi_\nu^2$ values. This region contains young O-stars and the gas flows around these stars likely violate the LTE assumptions or Gaussian line profiles. We see in the fits that the fitted $\tau$ values are either very low or very high (reflecting the degeneracy highlighted in Eq.\ \ref{eq:temp_sum}), and that often the fitted excitation temperatures will also be either very high or very low. We therefore suggest inspection of the corner plots (see Fig. \ref{fig:app_corner_plot} for an example) for fits where $\tau$ is at the prior bounds, as these are likely to be the regions where the LTE assumptions are inappropriate. These relatively poor fits, compared to the rest of the data, therefore reflect limitations in the simple profile for the line model, rather than a breakdown in the fitting routine.

Secondly, we investigate the number of components in Fig. \ref{fig:g316_75_n_comp}. The majority (55~per~cent) of our spectra are well-described by a single component, with a maximum of seven distinct components (0.02~per~cent of the fits). We find that the spectral complexity increases towards the central filament, as well as towards the star forming cores and clumps residing there. This can cause non-Gaussian line profiles \citep[e.g.][]{2020Peretto, 2024Rigby}. We leave the full interpretation to Watkins et al. (in preparation). Just under half of the fits required more than one component, indicating that this spectral complexity is ubiquitous throughout the filament, further motivating the need for automated, multi-component fitting. Despite there being no prior constraint of the number of components to fit, we find that the code typically prefers to fit a few components, rather than many weaker components.

To test any potential dependence on the quality of fits with the number of fitted components, we produce violin plots of $\chi_\nu^2$ against $n_{\rm comp}$ in Fig. \ref{fig:g316_75_n_comp_chisq}. There is a trend towards higher $\chi_\nu^2$ with increasing $n_{\rm comp}$. Reassuringly, this trend is \emph{opposite} to the idea that increasing the number of components may lead to overfitting (in which case, $\chi_\nu^2$ would systematically decrease with increasing $n_{\rm comp}$). In the vast majority of cases then, \codename\ produces reasonable fits to the data. We see some evidence of underfitting (i.e.\ potentially missed components in the spectrum), which we diagnose by manual inspection of the fits. These are either very narrow and weaker components, or very broad, very faint components -- in this case, adding an extra component does not substantially change the $\Delta {\rm BIC}$, and so \codename\ does not include this component. We prefer missing some components for the benefit of ensuring the fit components are trustworthy, although users can fine-tune this using the $\Delta {\rm BIC}$ if they desire.

\subsection{NGC~3627}\label{sec:fit_ngc3627}

Our second real-world benchmark is a nearby star-forming galaxy observed as part of the Physics at High Angular Resolution (PHANGS\footnote{\url{phangs.org}}) survey \citep{2021bLeroy}, NGC~3627 (project code ADS/JAO.ALMA\#2017.1.00886.L). The details of the data processing are given in \cite{2021aLeroy}, and are generally the same as G316.75, with the exception of the line, which is CO({\it J}=2-1) instead of \ntwohp, the cube is imaged at 2.5~km~s$^{-1}$ resolution, and {\tt feather} is used instead of {\tt sdintimaging}. NGC~3627 is a strongly barred galaxy with complex dynamics, particularly at the bar ends \citep{2017Beuther}, making this a good test for \codename. Using the PHANGS-ALMA data from the first public data release, we fit the cube in the same way as in Section \ref{sec:fit_g316}, although for the CO line there is no hyperfine splitting, simplifying the spectra considerably. Given that we only have a single component per spectrum fitted, this makes \tex\ and $\tau$ completely degenerate with each other, although the velocity and dispersion remain useful and well-constrained. Again, we fit around 100,000 pixels, and encouraging spatial coherence replaces around 1~per~cent of fits. As the CO line is only modelled as a single Gaussian, and we generally fit fewer components, the fitting here is faster than in Section \ref{sec:fit_g316}, taking around 40 hours.

We show the $\chi_\nu^2$ and $n_{\rm comp}$ maps in Figures \ref{fig:ngc3627_chisq} and \ref{fig:ngc3627_n_comp}. Unlike for G316.75, we see an extremely uniform distribution of $\chi_\nu^2$ values across the entire galactic disc, with a peak very close to one, and a very narrow spread of around 0.1 (16$^{\rm th}$ and 84$^{\rm th}$ percentiles). This shows the complexity of our model is sufficient to describe the line profiles in all $\sim$100~pc regions of these CO observations. The peak of the $\chi_\nu^2$ distribution is in the bar ends of this galaxy, which is the most dynamically active region with the highest level of spectral complexity. We also see that the majority of the CO lines-of-sight can be fit with a single component, but the spectral complexity increases towards the galaxy centre and the bar ends. Even in these relatively low-resolution CO observations, in dynamically complex regions, a tool like \codename\ is required to properly model the gas dynamics.

To highlight the importance of accounting for multiple velocity components along each line-of-sight, we show a comparison between the available PHANGS moment 1 (mom1; intensity-weighted average velocity) and effective width \citep[a proxy for the velocity dispersion, see][for a discussion of this quantity]{2021aLeroy} to our calculated values in Figure \ref{fig:mcfine_phangs_comparison}. For the PHANGS maps, we use the ``strict'' masked maps, which trades flux completeness for a lower false-positive rate. In the \codename\ case, when we have multiple components along the line-of-sight, we perform a peak temperature-weighted average of all the values, having removed any parameters that are not detected above a 5$\sigma$ limit, to reduce the noise. We also only show sightlines where we fit multiple components, as this is where we expect to see the majority of the deviations between the two methods. We can see that the velocity values are relatively similar, as expected as the mom1 map is also intensity averaged. However, the effective widths are significantly biased along sightlines with higher spectral complexity, with the effective width being significantly higher than the average velocity dispersion \codename\ calculates. This is because the effective width blindly takes all emission along a sightline into account, leading to extremely large effective widths for regions with multiple components more widely spaced in velocity. In dynamically complex regions, it is therefore crucial to account for this spectral complexity. Henshaw et al. (in preparation) perform a similar comparison across the full PHANGS sample, finding very comparable results.

As one further test, we compare \codename\ to another multi-component fitting code. In Figure \ref{fig:mcfine_scouse_comparison}, we show a comparison of the number of components fitting by \codename\ to those found by {\tt scouse}, a semi-automated multi-component line fitting technique \citep{2016Henshaw, 2019Henshaw}, which are presented in Henshaw et al. (in preparation). In the majority of cases, \codename\ and {\tt scouse} fit the same number of components. In around 30~per~cent of cases, {\tt scouse} fits more components than \codename, but the opposite is much less frequent (3~per~cent). In general, these two codes suggest very similar levels of spectral complexity of NGC~3627.

\section{Conclusions}\label{sec:conclusions}

In this work, we have presented \codename, an MCMC tool for fitting complex spectral profiles in astronomical data. \codename\ works by building up steadily more complex models, and performing comparisons using the Bayesian Information Criterion to decide an acceptable stopping point for the complexity of the model. In the case of data cubes (e.g. from ALMA), it can also encourage spatial coherence by comparing neighbouring models. In a number of synthetic tests, we have shown that \codename\ overall is successful at recovering our input parameters, fitting the correct number of components in around 90~per~cent of cases. We tend to recover the correct parameters (within the fitting uncertainties) for spectra where we fit the correct number of components. 
For fits with too few components, either the lines are highly blended and confused with a single brighter line, or the line is faint enough that it did not strongly affect the combined spectrum.
Our fits well-describe the data, with a median $\chi_\nu^2$ value of 1.01 across our synthetic tests, and a small spread of around 0.1 (16$^{\rm th}$and 84$^{\rm th}$ percentiles).

We also tested \codename\ on two large ALMA mosaics as a real-world example. For the first, we use LTE modelling to fit $\simeq$100,000 \ntwohp\ spectra from a Galactic molecular filament. \codename\ most commonly fits a single component, with a maximum of seven components (0.02~per~cent of the fits). There are a significant number of spaxels (45~per~cent) that require more than one component. The $\chi_\nu^2$ distribution is very close to a value of one, indicating that on average the fits almost perfectly describe the data. However, there are some regions that have elevated $\chi_\nu^2$ values. These are complex, active regions of star formation where the LTE or Gaussian line assumptions are likely to break down due to the complex geometry and motions of the gas. These regions have either very high or low $\tau$ values, and are likely where full radiative transfer modelling is required. We find a slight increase in the $\chi_\nu^2$ values with the number of fitted components, indicating that this spectral complexity is needed to help properly model the gas in this region, rather than \codename\ overfitting the data.

Although \codename\ has primarily been developed to work with \ntwohp\ emission, it is relatively simple to extend to other lines via simple dictionaries within the code. In our second test, we demonstrate this adaptability by fitting an ALMA CO({\it J}=2-1) mosaic of NGC~3627, a nearby star-forming galaxy. We find a smooth $\chi_\nu^2$ distribution across all of our fitted spectra, showing that the model successfully captures the variations in the data. The spectral complexity increases towards the ends of the bar and the galaxy centre, as we would expect in dynamically complex regions. We have shown that accounting for these multiple components is critical to obtain reasonable measures of the velocity dispersion of the gas. In comparison to another multi-component fitting code, {\tt scouse}, we generally fit the same number of components, with {\tt scouse} fitting more components in around 30~per~cent of spaxels, but with an overall similar $\chi_\nu^2$ value. \codename\ fits more components than {\tt scouse} in around 5~per~cent of cases. This comparison gives us confidence that our methodology does not lead to wildly different results to other fitting philosophies.

As radio telescopes such as ALMA and NOEMA push towards large mosaics of thousands of independent points covering multiple diagnostic emission lines, tools like \codename\ are required to interpret these rich datasets, rather than a more bespoke, `hands-on' approach. As such, an open-source tool like \codename\ will be invaluable for the community to use in their science.

\section*{Acknowledgements}

We thank the reviewer for their comments which have improved the quality and presentation of this work. The authors would also like to thank Michael Anderson, Andrew Rigby, and Nicolas Peretto for useful discussions, and to Jonathan Henshaw for sharing the {\tt scouse} fits for NGC~3627. EJW acknowledges the support of STFC consolidated grant number ST/X001229/1. 

This paper makes use of the following ALMA data: ADS/JAO.ALMA\#2013.1.01029.S, ADS/JAO.ALMA\#2018.1.01851.S, ADS/JAO.ALMA\#2017.1.00886.L. ALMA is a partnership of ESO (representing its member states), NSF (USA) and NINS (Japan), together with NRC (Canada), MOST and ASIAA (Taiwan), and KASI (Republic of Korea), in cooperation with the Republic of Chile. The Joint ALMA Observatory is operated by ESO, AUI/NRAO and NAOJ.

\section*{Data Availability}

The source code for \codename\ is available at \url{https://github.com/thomaswilliamsastro/mcfine}, and the testing scripts at \url{https://github.com/thomaswilliamsastro/mcfine_testing}. \codename\ can be installed with {\tt pip install mcfine}, although the RT capabilities will require {\tt gfortran} to be in the user PATH. The PHANGS-ALMA data is publicly available online at \url{https://www.canfar.net/storage/list/phangs/RELEASES/PHANGS-ALMA/}. The G316.75 data will be presented in Watkins et al. (in preparation).



\bibliographystyle{mnras}
\bibliography{bibliography}



\appendix

\section{Example Synthetic Fits}\label{app:synth_fits}

Here, we give the fits for the lowest, median, and highest $\chi_\nu^2$ fits from our synthetic tests in Sect. \ref{sec:synth_tests}. In the case for the model with the worst $\chi_\nu^2$, we see that \codename\ fits fewer components than the input spectrum, leading to a worse fit to the overall spectrum.

\begin{figure*}
\includegraphics[width=\textwidth]{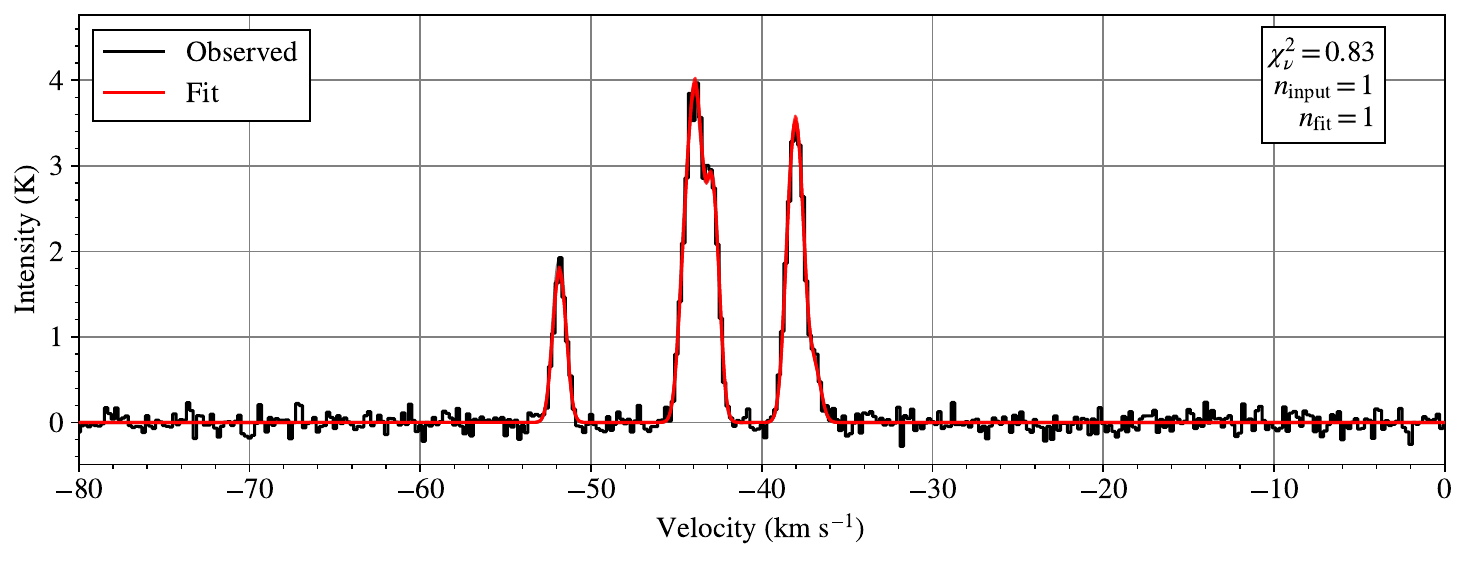}
\caption{Fit spectrum for the synthetic model with the lowest $\chi_\nu^2$ in our tests. The red line shows the fit to the data, shaded red region the fit uncertainty. In the upper-right corner, we give the $\chi_\nu^2$ value, and the input and fitted number of components.} 
\label{fig:app_lowest_chisq}
\end{figure*}

\begin{figure*}
\includegraphics[width=\textwidth]{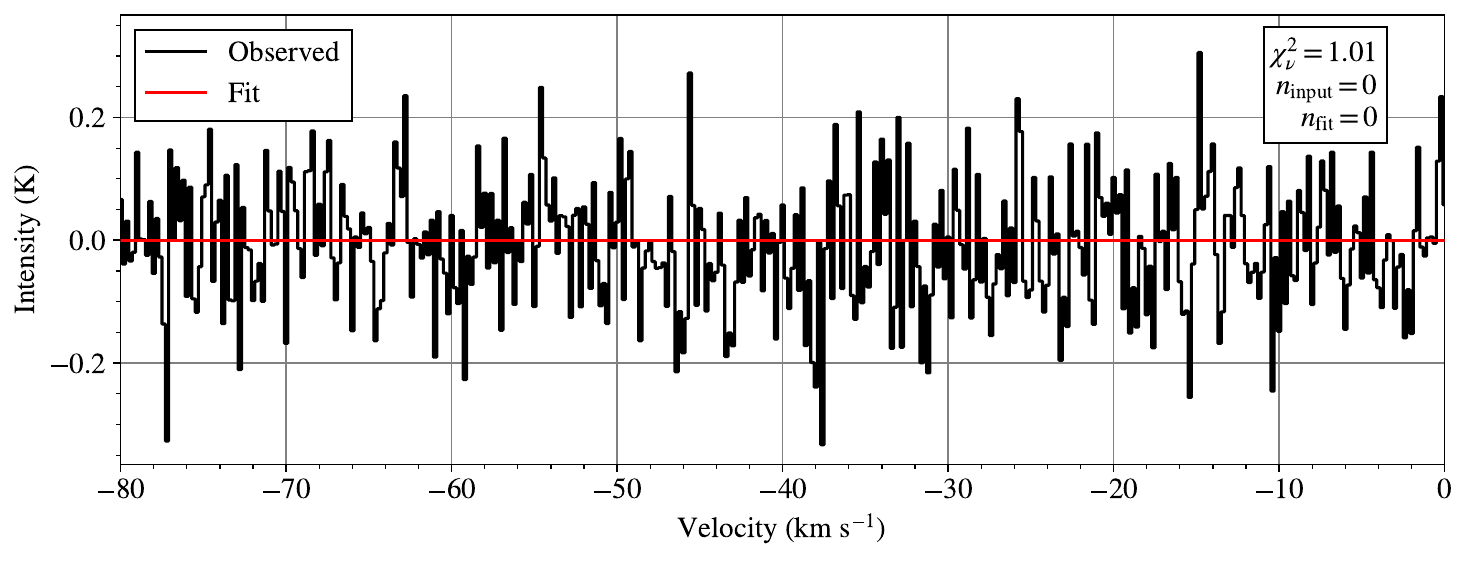}
\caption{As Fig. \ref{fig:app_lowest_chisq}, but for the fit with the median $\chi_\nu^2$.} 
\label{fig:app_med_chisq}
\end{figure*}

\begin{figure*}
\includegraphics[width=\textwidth]{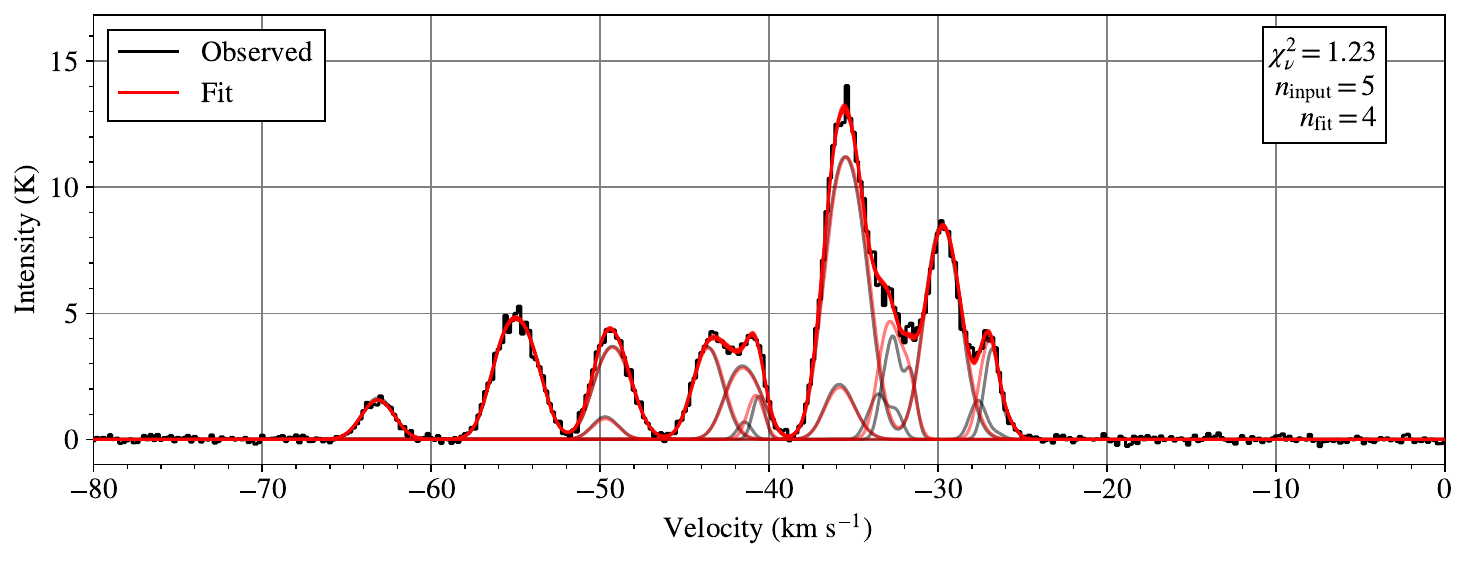}
\caption{As Fig. \ref{fig:app_lowest_chisq}, but for the fit with the worst $\chi_\nu^2$. The faint black lines show the individual components, and the faint red lines the fit to them. In this case, we fit one fewer component than input, but this is due to there being two relatively faint, blended components.}
\label{fig:app_high_chisq}
\end{figure*}

\section{Example ALMA Fits}\label{app:alma_fits}

Here, we give the fits for the lowest, median, and highest $\chi_\nu^2$ fits from our real-world tests to G316.75 in Sect. \ref{sec:real_app}. For the fit with the median $\chi_\nu^2$, we also show the corner plot.

\begin{figure*}
\includegraphics[width=\textwidth]{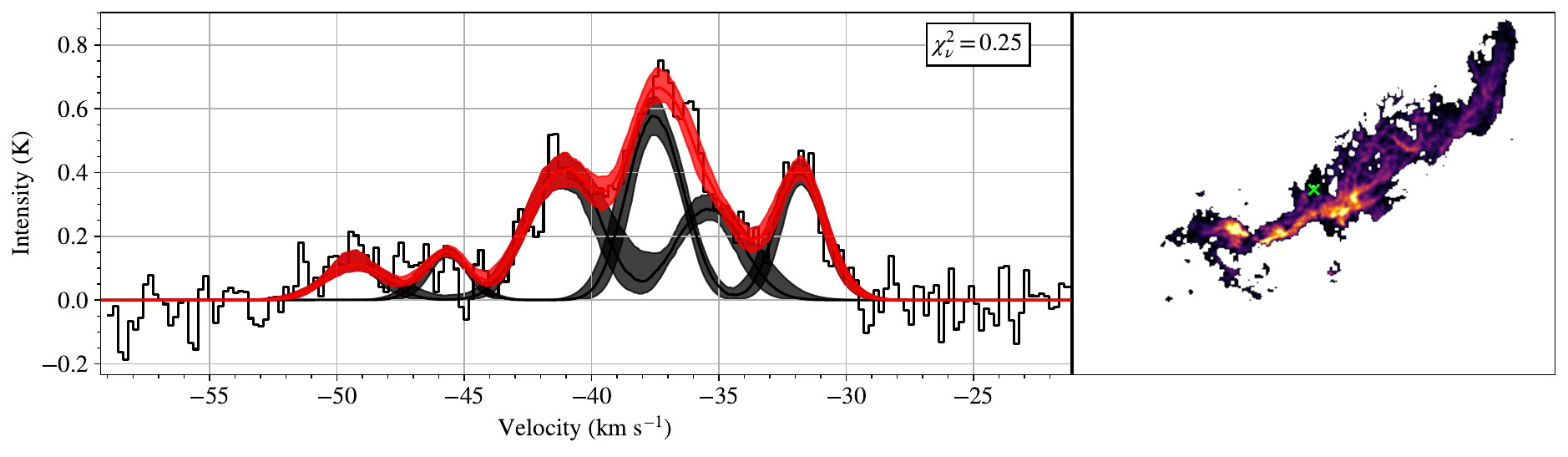}
\caption{{\it Left:} Fit spectrum for the model with the lowest $\chi_\nu^2$ in our ALMA data. The red line shows the fit to the data, shaded red region the fit uncertainty. We also show the fit and uncertainty for each individual component. In the upper-right corner, we give the $\chi_\nu^2$ value. {\it Right}: Integrated intensity map for the region, with the location of this spaxel shown as a bright green cross.} 
\label{fig:app_lowest_chisq_alma}
\end{figure*}

\begin{figure*}
\includegraphics[width=\textwidth]{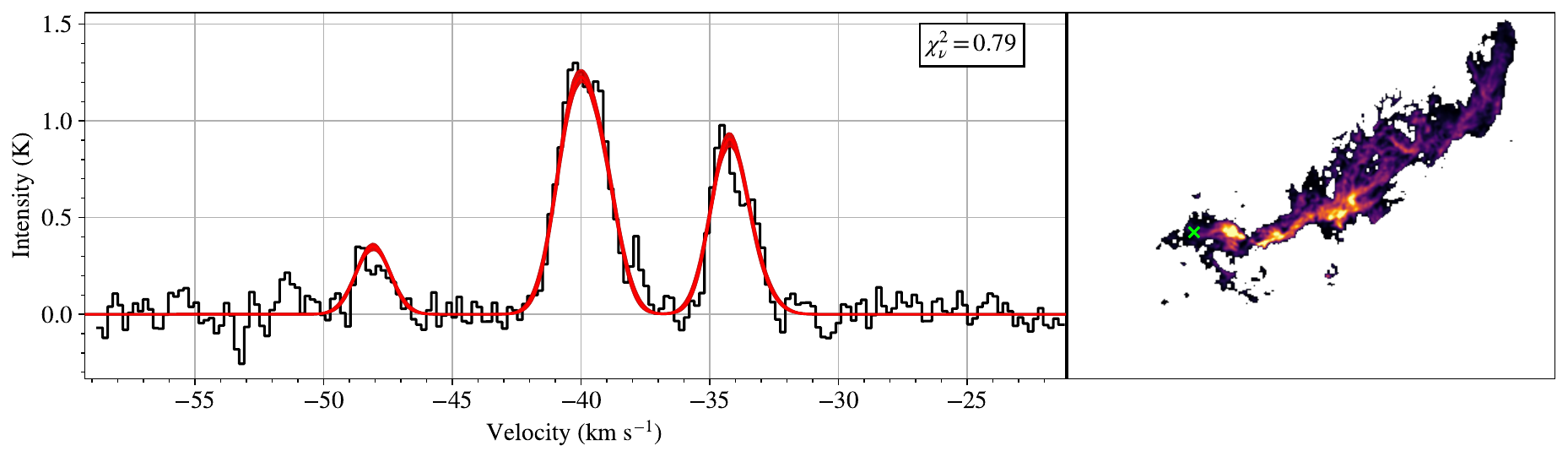}
\caption{As Fig. \ref{fig:app_lowest_chisq_alma}, but for the fit with the median $\chi_\nu^2$.} 
\label{fig:app_corner_plot}
\end{figure*}

\begin{figure*}
\includegraphics[width=.75\textwidth]{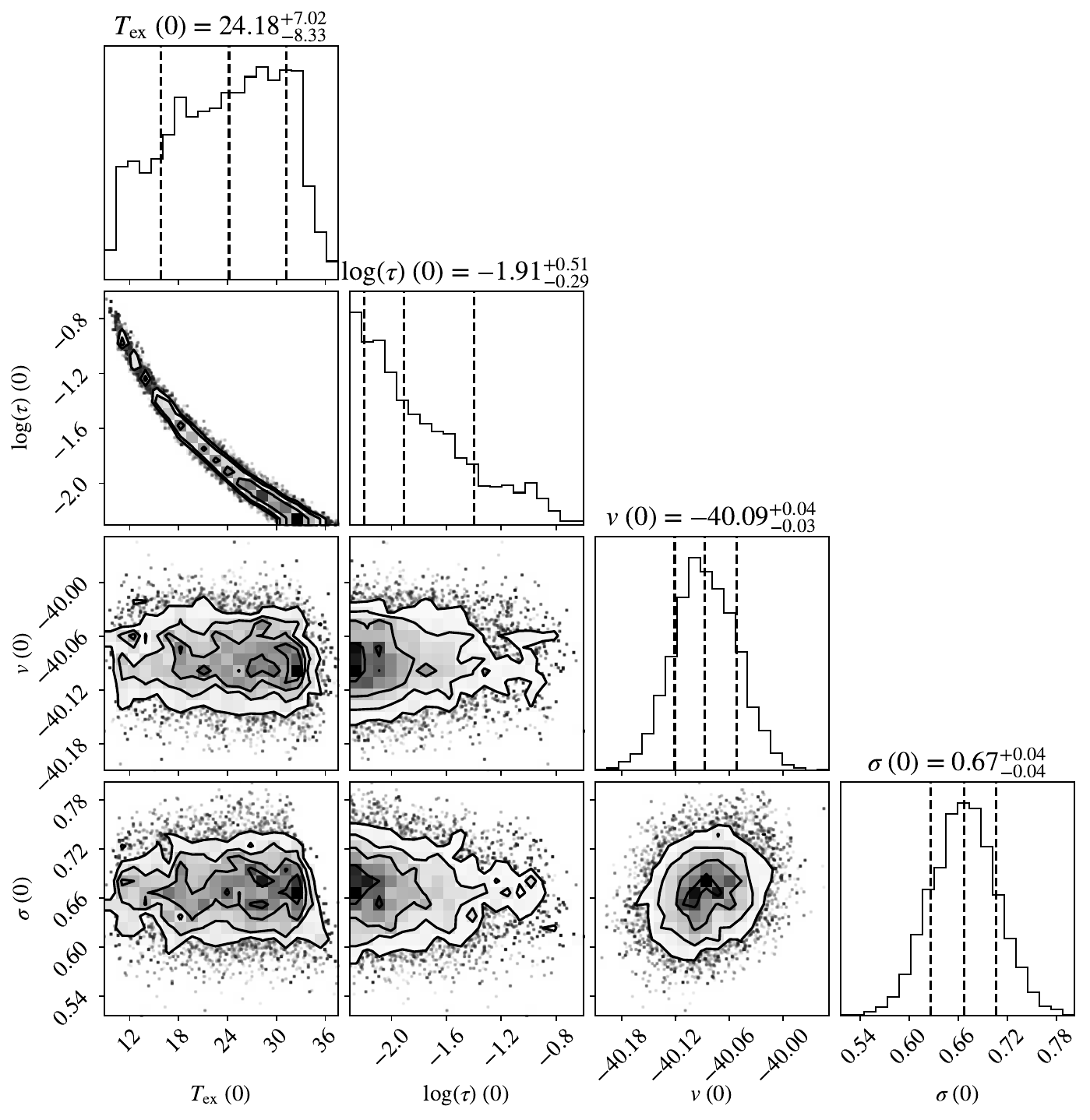}
\caption{Corner plot for the fit shown in Figure \ref{fig:app_med_chisq_alma}.} 
\label{fig:app_med_chisq_alma}
\end{figure*}

\begin{figure*}
\includegraphics[width=\textwidth]{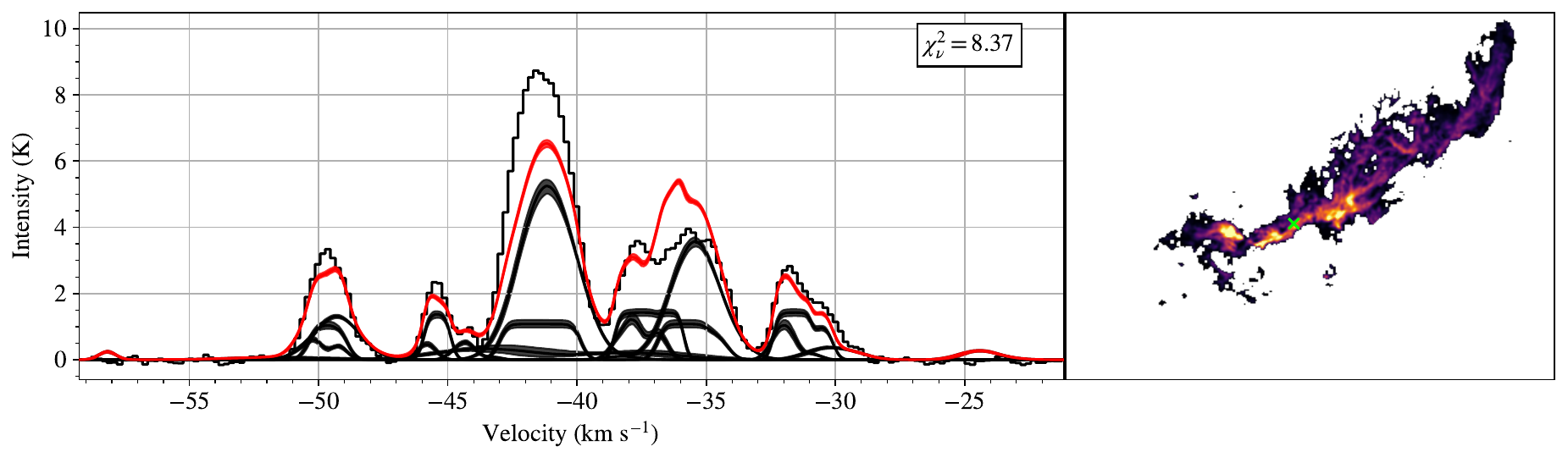}
\caption{As Fig. \ref{fig:app_lowest_chisq_alma}, but for the fit with the highest $\chi_\nu^2$.}
\label{fig:app_high_chisq_alma}
\end{figure*}


\bsp	
\label{lastpage}
\end{document}